\documentclass[%
 reprint,
 amsmath,amssymb,
 aps,
]{revtex4-2}

\usepackage{graphicx}
\usepackage{dcolumn}
\usepackage{bm}
\usepackage{float}
\usepackage{enumerate}
\usepackage{xcolor}

\begin{document}

\preprint{APS/123-QED}

\title{Families of Discrete Breathers on a Nonlinear Kagome Lattice}

\author{Andrew Hofstrand}
 \affiliation{Mathematics Department, New York Institute of Technology, New York, NY 10023}
 \email{ahofstra@nyit.edu}

\date{\today}

\begin{abstract}
The unique geometry of the two-dimensional tripartite Kagome lattice is responsible for shaping diverse families of spatially localized and time-periodic nonlinear modes known as discrete breathers.  We state conditions for the existence of breathers and compute their spatiotemporal profiles near the edges of the linear phonon spectrum. Our findings include the existence of strongly nonlinear and dynamically stable breathers inside the band gap on the infinite lattice, asymptotic expressions for breather energy thresholds in the weakly nonlinear regime, and explicit breather solutions that remain compactly supported on the lattice and undergo stability transitions.
\end{abstract}

\maketitle

\section{Introduction}
The study of nonlinear waves and coherent structures on both classical and quantum lattices has a long and venerable history with diverse applications in condensed matter physics~\cite{kivshar,ober}, optical science~\cite{moloney,turitsyn}, mechanical metamaterials~\cite{hecke,chen}, and circuit theory~\cite{alu}.  Theoretical aspects, such as the stability and localization properties of nonlinear modes, play an important role in the design of novel platforms~\cite{li}. A universal class of solutions on nonlinear lattices known as \textit{discrete breathers}~\cite{aubry}, which are periodic in time and spatially localized, are vital to understanding lattice dynamics with broad classes of initial data.  Indeed, both transient and long-time asymptotic behavior, including energy trapping in metastable states and the transition to thermodynamic equilibrium, are intimately connected to the existence of discrete breathers on the lattice~\cite{zabusky,xiong1,xiong2,wayne}. The properties of discrete breathers in both conservative and damped-driven systems have been studied extensively~\cite{flach}, and include existence proofs, minimum energy thresholds, numerical continuation schemes, stability analyses, and the phenomena of nanoptera~\cite{porter}. Unlike their continuous counterparts, the fact that the linear spectrum of the lattice is bounded makes large families of discrete breathers possible in any spatial dimension.

Two-dimensional lattices are of particular interest as toy models in physical systems due to advances in mono-layer crystal fabrication in material science\cite{liu}, along with applications such as trapping ultracold atoms in optical lattices~\cite{kurn0}.  Many aspects of including nonlinearity in these models are still unexplored and present novel opportunities for controlling energy localization.  The Kagome lattice with its tripartite structure has recently attracted attention due to its distinct geometry~\cite{hasan,vitelli,khan}, which gives rise to a band structure with a strictly flat band, touching one of the other dispersive bands at the $\Gamma$-point, and Dirac cones located at the corners of the hexagonal Brillouin zone (see Figure \ref{figure1}).  The topology of the Kagome lattice is responsible for diverse phenomena such as frustrated spin dynamics in tight-binding models of magnetic systems~\cite{heyderman}, spatially compact flat-band eigenstates~\cite{balents}, and the existence of zero-energy modes that are pinned to the corners of a triangular lattice and exhibit a generalized chiral symmetry~\cite{khanikaev,berc,buljan}.  Notably, because of destructive interference effects on the Kagome lattice, the latter modes remain localized even when energetically resonating with the lattice's phonon spectrum.  We consider here a well-known generalization of the Kagome lattice called the breathing Kagome (BK) lattice~\cite{ezawa,kurn,kempkes}, where the \textit{intra}-unit-cell and \textit{inter}-unit-cell coupling strengths can be unequal, reducing the lattice's point symmetry group from $C_6$ to $C_3$ and opening a band gap at the high-symmetry points.  We remark that the term ``breathing" here has nothing to do with discrete breathers.  There have been a few recent works on \textit{nonlinear} dynamics in Kagome lattices, including a study of vortex breathers~\cite{kev2}, a stability analysis of spatially compact breathers bifurcating from flat-band eigenstates~\cite{johansson}, and investigations on nonlinear corner and edge modes~\cite{heinrich,theo}.  The results in~\cite{kev2,johansson} concern breathers in the discrete nonlinear Schr\"odinger (dNLS) equation, i.e. standing waves (or ``trivial breathers"), rather than genuine periodic orbits which are considered here. 

In the current work, we state conditions for the existence of discrete breathers on a nonlinear BK lattice and iteratively compute their spatiotemporal profiles, seeded from the so-called \textit{molecular limit}~\cite{hof,benal}--a simple generalization of the anti-continuum limit first introduced in~\cite{aubry}.  Importantly, when continuing from the molecular limit, the unequal but spatially symmetric coupling strengths provide a means to obtain discrete breathers with frequencies located inside band gaps.  After finding exact periodic solutions to the equations of motion on an isolated unit cell, we find breathers on the connected lattice by tuning the inter-cell coupling strength from zero up to a value where the \textit{fixed} breather frequency is located either just above the top band, below the bottom band, or inside the band gap of the linear phonon spectrum.  Due to the diversity of features in the Kagome lattice's dispersion relations, the continued breathers have very different localization properties, dependent on the curvature of the nearest phonon band edges.  We first construct discrete breathers inside the small phonon gap, where the band edges approach Dirac-crossings. We show that a family of strongly nonlinear and dynamically stable soliton-like breathers exist on the infinite lattice, exhibiting a rotational sub-lattice symmetry. Next, we derive an asymptotic long-wave model for weakly localized breathers bifurcating from the lattice's bottom parabolic band edge. Our asymptotics provide analytic expressions for breather norms and indicate accurately how breathers terminate at the band edge.  Finally, we provide exact breather solutions that take nonzero values only on a closed ring within the extended lattice.  We compute the stability transitions of these breathers as a function of coupling strength and lattice size and state possible directions for future investigation in the conclusion.
\section{Results}
\subsection{Model Description}
\begin{figure}[h]
    \includegraphics[scale=0.52]{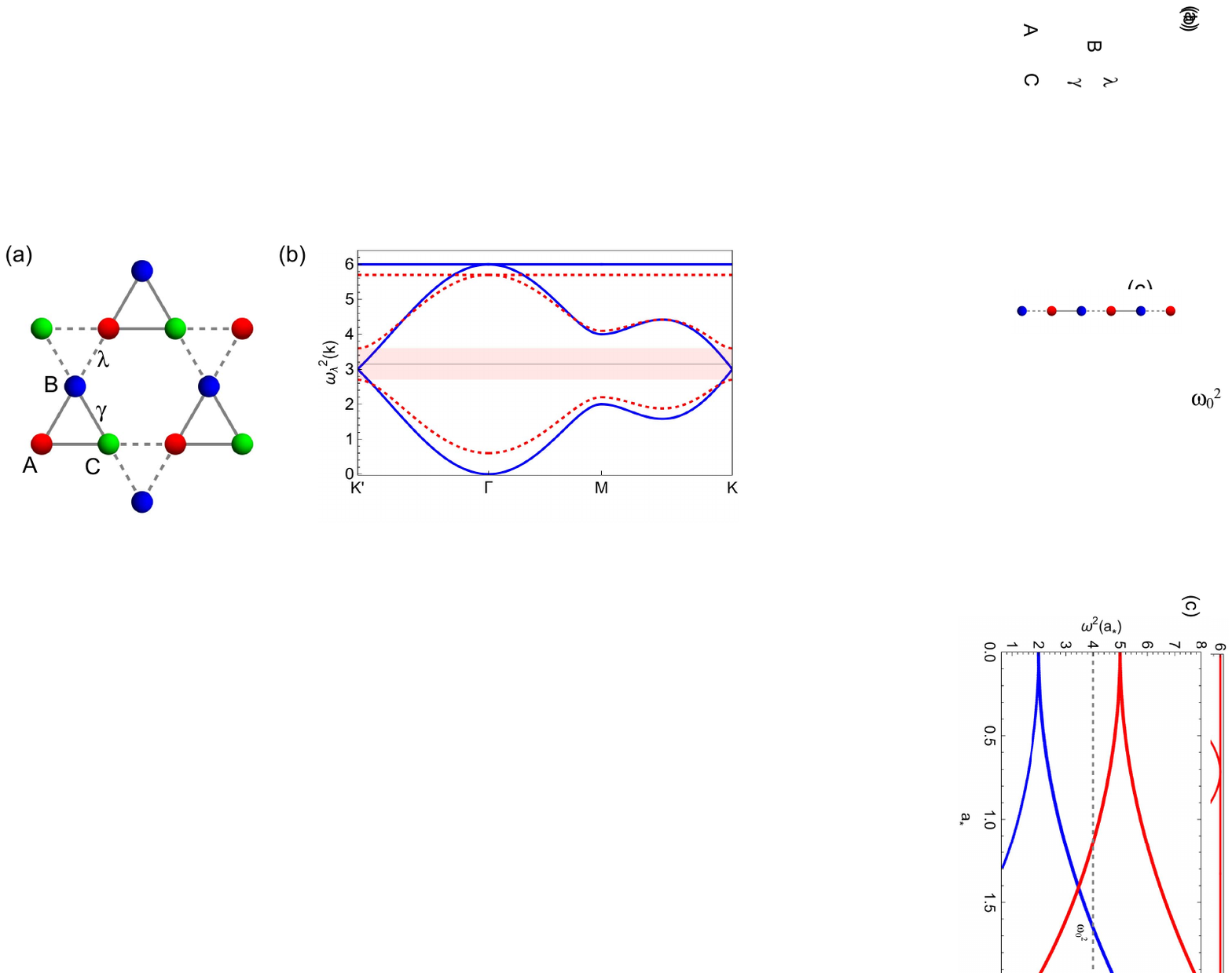}
    \caption{(a) schematic of the BK lattice in \eqref{eq1}; (b) phonon bands of BK lattice ($\omega_0=2$, $\gamma=1$) when $\lambda=1$ (blue) and $\lambda=0.7$ (red, dashed); the band gap is shaded.}
    \label{figure1}
\end{figure}

We study a nonlinear mass-spring BK lattice, depicted in Figure \ref{figure1}(a). The BK lattice consists of a 3-site unit-length equilateral triangular cell (A,B,C) and tiles the plane with lattice vectors $\mathbf{a}_1=(2,0)$ and $\mathbf{a}_2=(1,\sqrt{3})$, indexed by $\alpha:=(n,m)\in\mathbb{Z}^2$.  Considering linear nearest-neighbor interactions, the equations of motion on the infinite lattice are 
\begin{equation}\label{eq1}
\begin{aligned}
&\ddot{x}_{\alpha}^A=-V'(x_{\alpha}^A)+\gamma\left[x_{\alpha}^B+x_{\alpha}^C\right]+\lambda\left[x_{n,m-1}^B+x_{n-1,m}^C\right]\\
&\ddot{x}_{\alpha}^B=-V'(x_{\alpha}^B)+\gamma\left[x_{\alpha}^A+x_{\alpha}^C\right]+\lambda\left[x_{n,m+1}^A+x_{n-1,m+1}^C\right]\\
&\ddot{x}_{\alpha}^C=-V'(x_{\alpha}^C)+\gamma\left[x_{\alpha}^A+x_{\alpha}^B\right]+\lambda\left[x_{n+1,m}^A+x_{n+1,m-1}^B\right].
\end{aligned}
\end{equation}

We take the on-site nonlinear potential $V$ to be even and given by
\begin{equation}\label{eq2}
V(z)=\dfrac{1}{2}\omega_0^2z^2+\dfrac{g}{2\sigma+2} |z|^{2\sigma}z^2,
\end{equation}
where $\sigma\geq 1$ and $g=\pm 1$ for a \textit{hardening} or \textit{softening} nonlinearity, respectively.  Here, the in-cell coupling $\gamma$ is fixed at some positive value and the out-of-cell coupling $\lambda$ is variable. The derivation of the linear lattice's band structure is outlined in Appendix A and is represented in Figure \ref{figure1}(b).

Notably, the top-most phonon band is flat over the entire Brillouin zone and touches the maximum of the next-lowest dispersive band at the $\Gamma$-point regardless of the values of the coupling parameters. When $\lambda=\gamma$ the band structure has Dirac cones at the quasi-momentum $K$ and $K'$ points.  A band gap opens at these points when $\lambda\neq\gamma$ as shown in Figure \ref{figure1}(b).  The width of the gap is $3|\gamma-\lambda|$ and the center of the gap is $\omega_0^2-(\gamma+\lambda)/2$.  Note that, since the matrix resulting from the linearization of system \eqref{eq1} about the zero solution has a constant trace, the three eigenvalues $\omega^2_{1,2,3}$ at any fixed quasi-momentum $\mathbf{k}$ must add to $3\omega_0^2$ (see Appendix A).
\subsection{Existence of Breathers}

The existence of discrete breathers on the globally coupled BK lattice can be proven near the molecular limit, defined as taking $\lambda=0$ in \eqref{eq1}, i.e. when each unit cell is disconnected from the others.  Indeed, given a nonzero, even $T_b$-periodic solution at the molecular limit, an application of the implicit function theorem proves the existence of a $\lambda_b>0$ and discrete breathers on the interconnected BK lattice, for $0<\lambda<\lambda_b$, given that the following two mild conditions hold: 
\begin{enumerate}
\item\underline{Non-resonance}: For a breather with frequency $\omega_b:=2\pi/T_b$, we have $(n\omega_b)^2\neq V''(0)+\gamma$ and $(n\omega_b)^2\neq V''(0)-2\gamma$ for all $n\in\mathbb{Z}$.
\item\underline{Non-degeneracy}: This condition is more nuanced, but holds quite broadly given that the seeded solution is even-in-time and that the derivative of the breather's period with respect to its energy is non-vanishing, $\partial{T_b(E)}/\partial{E}\big|_{E_*}\neq 0$ (see \cite{aubry, hof} for a detailed discussion). 
\end{enumerate}
Satisfying the preceding conditions implies the \textit{existence} of a $\lambda_b>0$, but not how to obtain its largest value.  In what follows, we numerically construct breathers by expressing \eqref{eq1} in terms of its temporal Fourier coefficients and using a Newton-Raphson iteration scheme, seeded by a known periodic solution in the molecular limit~\cite{hof}.  Our numerics suggest that breathers persist until ``nearly" intersecting the linear lattice's phonon spectrum, which deforms with $\lambda$ (see Appendix A).  Throughout this short paper, we denote the out-of-cell coupling where $\omega_b$ first intersects the phonon spectrum by $\lambda_*$. Typically, once the breather frequency (or one of its higher harmonics) intersects the phonon spectrum, breathers will couple to radiating plane waves, losing localization in conservative systems like \eqref{eq1}~\cite{flach,hof}.  However, in certain lattices, such as the BK lattice with its flat dispersion band, compactly localized breathers exist for $\lambda\neq 0$ (see below) and, due to destructive interference effects, cannot radiate energy via plane waves away to infinity even if $\omega_b$ intersects the phonon bands. We find that compact breathers on the BK lattice are spectrally unstable when $\omega_b$ is located inside the phonon bands. 
 
\begin{figure*}[t!]
\centering 
    \includegraphics[width=16cm,height=16cm,keepaspectratio,angle=90]{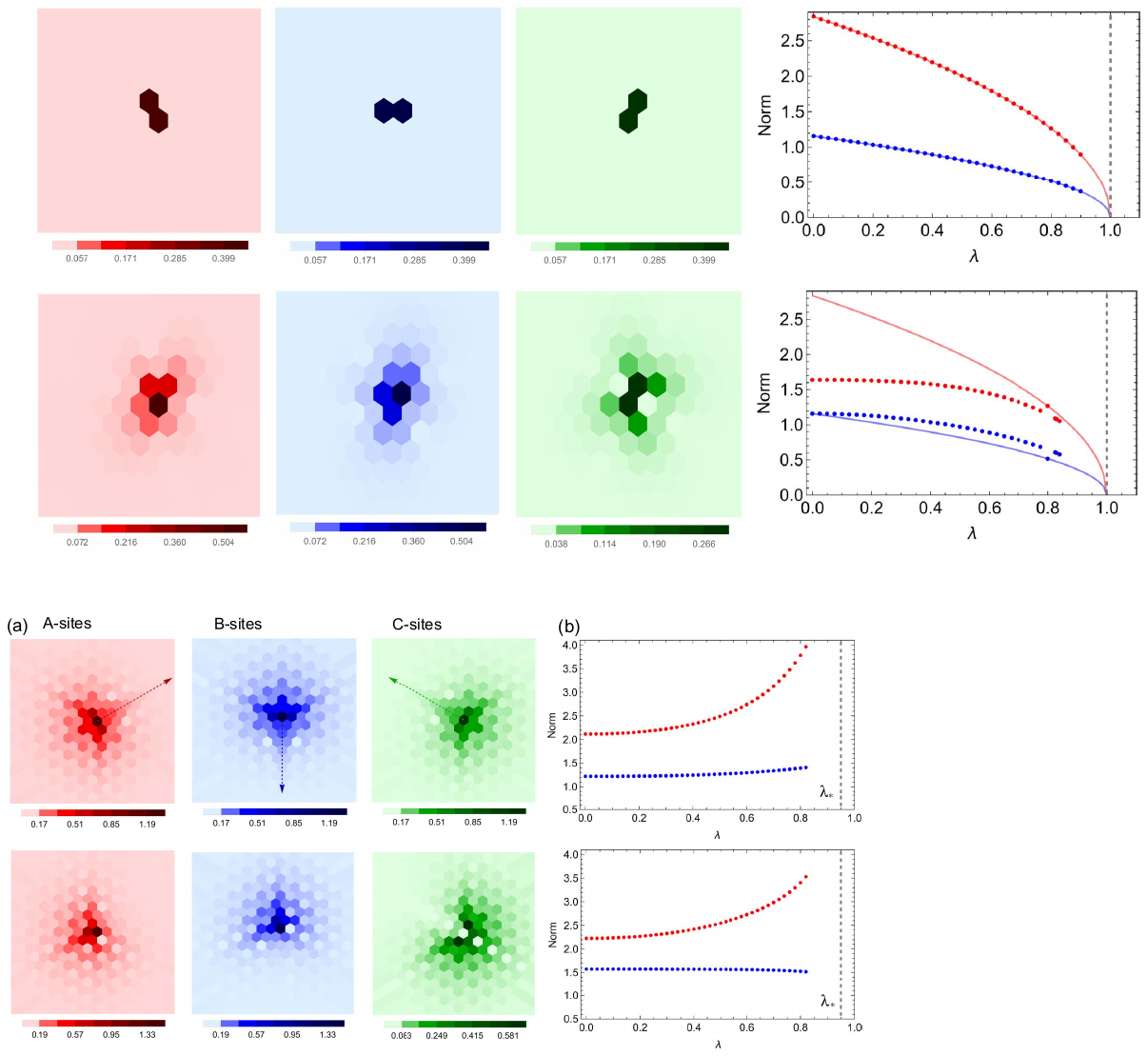}
    \caption{\textbf{Top row of panels}: breathers 
    with $\omega_b$ inside the phonon gap, seeded from a \textit{in-phase} state with $g>0$ (a) the spatial profile (absolute value) of the mid-gap breather when $\lambda=0.8$ at $t=0$, projected separately on its three sub-lattices (arrows highlight $C_3$ sub-lattice symmetry); (b) infinity- (blue) and 2- (red) norms of the breather versus inter-cell coupling ($\lambda_*=0.95$ dashed vertical line); \textbf{Bottom row of panels}: same as (a) but seeded from an \textit{out-of-phase} state with $g<0$. (parameters: $\gamma,|g|=1$, $\omega_0=2$, $\omega_b^2=3.1$, ).}
    \label{figure2}
\end{figure*}
\subsection{Molecular Limit States}

The molecular limit of \eqref{eq1} is a dynamical system with a six-dimensional phase space that admits families of quasi-periodic and possible chaotic trajectories.  Here, we effectively reduce the dimension of the phase space to two by seeding the lattice with the following solutions in the molecular limit: 
\begin{enumerate}[I]
\item \underline{In-phase states}: Given $x_*^{A,B,C}(0)=a_*\neq 0$ and $\dot{x}_*^{A,B,C}(0)=0$, there exists a periodic solution
\[\begin{pmatrix}x_*^A(t)\\x_*^B(t)\\x_*^C(t)
\end{pmatrix}=z_*^{(I)}(t)\begin{pmatrix} 1\\1\\1
\end{pmatrix}\]
so long as $z_*^{(I)}$ is a periodic solution of the reduced dynamical system
\[\ddot{z}=-V'(z)+2\gamma z,\quad z(0)=a_*,\quad \dot{z}(0)=0.\]
\item \underline{Out-of-phase states}: Given $x_*^{A,B,C}(0)=a_*(1,-1,0)^{\top}$ or any permutation thereof and $\dot{x}_*^{A,B,C}(0)=0$, there exists a periodic solution
\[\begin{pmatrix}x_*^A(t)\\x_*^B(t)\\x_*^C(t)
\end{pmatrix}=z_*^{(II)}(t)\begin{pmatrix} 1\\-1\\0
\end{pmatrix}\]
so long as $z_*^{(II)}$ is a periodic solution of the reduced dynamical system
\begin{equation}\label{out}
    \ddot{z}=-V'(z)-\gamma z,\quad z(0)=a_*,\quad \dot{z}(0)=0.
    \end{equation}
\end{enumerate}
Given the form of $V$, there are in- and out-of-phase periodic states, depending on their initial amplitude $a_*$.  
\subsection{Breather Families}
\subsubsection{Gap Breathers}
In Figure \ref{figure2} we consider two breather families, each having the same frequency, $\omega_b$, located within the phonon band gap ($0\leq\lambda<\lambda_*$).  For the chosen parameters, when $\lambda=0.8$, $\omega_b$ is located exactly at the center of the small phonon band gap. We seed the lattice's central unit-cell with either an in-phase state for the case of a hardening nonlinearity or an out-of-phase state for the case of a softening nonlinearity.  The respective results are shown in the top (bottom) panels of Figure \ref{figure2}.  In the case of an in-phase seed, the breather's $C_3$ sub-lattice symmetry persists onto the globally coupled lattice, as seen in the top row of Figure \ref{figure2}(a). The spatial profile has been projected onto its three sub-lattice components to clearly observe this symmetry.  The out-of-phase breather remains rotationally asymmetric, as seen in the bottom panels of Figure \ref{figure2}(a) when $\lambda=0.8$. It has recently been noted~\cite{khanikaev, buljan, berc} that the linear BK lattice exhibits a generalized chiral or sub-lattice symmetry.   Remarkably, the symmetric in-phase breathers are \textit{dynamically stable} on the infinite lattice and both the breather amplitude and its $\ell^2$- or energy norm grow near the band edges, (see the norms in Figure \ref{figure2}(b) versus continuation parameter $\lambda$).  These strongly nonlinear and localized states should be contrasted with small-amplitude discrete breathers due to tangent bifurcations from band-edge plane waves~\cite{flach3}.  

Due to the Hamiltonian structure of \eqref{eq1}, breathers are dynamically stable if and only if the eigenvalues of the so-called \textit{monodromy matrix}, computed from the solution to the linearized $T_b$-periodic equations about the given breather, lie along the unit circle in the complex plane (Floquet multipliers)~\cite{promislow,flach}. 
\begin{figure}[H]
    \includegraphics[scale=0.61]{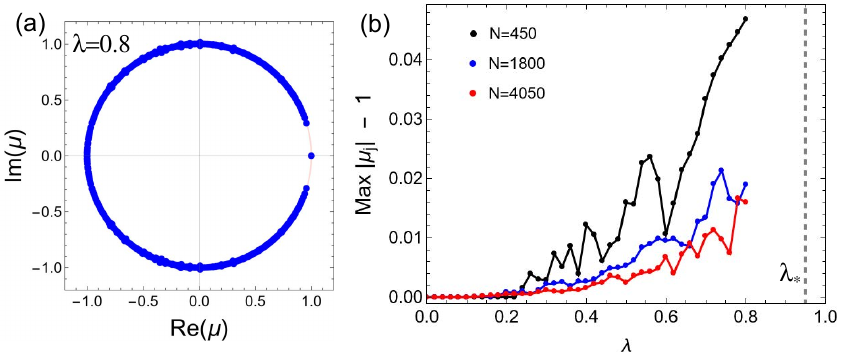}
    \caption{(a) Floquet spectrum of the breather shown in the top panels of Figure \ref{figure2}(a); (b) maximum deviation of the Floquet spectrum's modulus from one versus $\lambda$ for the breathers shown in the upper panel of \ref{figure2}(b) for the total lattice sizes $N=450,1800,4050$ sites.}
    \label{figure3}
\end{figure}
\noindent The Floquet spectrum of the in-phase mid-gap breather at $\lambda=0.8$ is shown in Figure \ref{figure3}(a).  We remark that the Floquet exponent at $+1$ is due to the discrete time-translational invariance of the breather and remains fixed for all $0\leq\lambda<\lambda_b$. The slight deviations of the spectrum from the upper portion of the unit circle in Figure \ref{figure3}(a), due to colliding Floquet exponents, result from the finite size of the lattice~\cite{marin}. Indeed, in Figure \ref{figure3}(b) we track the departure of the maximum modulus of the Floquet exponents from one as a function of $\lambda$ for different lattice sizes.  As $N$ increases the deviations become less pronounced and on the infinite lattice go to zero~\cite{marin}.  In contrast, the asymmetric breathers in the bottom panels of Figure \ref{figure2} are all dynamically unstable. For each $\lambda$, the instability arises due to the existence of a single Floquet multiplier with a value much greater than one on the real axis and a corresponding localized eigenvector (not shown).

\subsubsection{Breathers Beneath the Acoustic Band}
\begin{figure}[h]
\hspace*{-0.3cm}
    \includegraphics[width=0.5\textwidth,keepaspectratio]{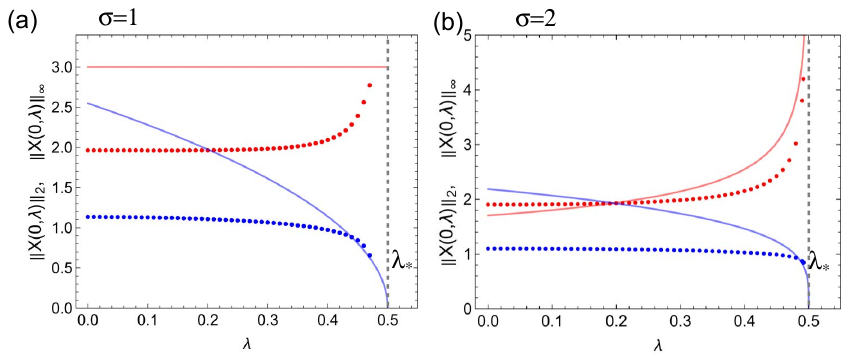}
    \caption{Infinity- (blue dots) and 2- (red dots) norms of breather family with $\omega_b$ located below the bottom phonon band, continued from a symmetric state in the molecular limit for $g<0$ when (a) $\sigma=1$ and (b) $\sigma=2$.  Solid curves indicate the corresponding asymptotic norms calculated from the expressions given in the text, showing agreement as $\lambda\to\lambda_*$. (parameters: $g=-1$, $\omega_b=1$, $\lambda_*=0.5$).}
    \vspace*{-0.5cm}
    \label{figure4}
\end{figure}

Next, we consider a family of breathers that have a frequency located underneath the BK lattice's lowest phonon band.  As $\lambda$ increases, the parabolic band edge at the $\Gamma$-point approaches $\omega_b$ from above.  To analyze this situation, we introduce the parameter $\epsilon:=\gamma/2-\lambda$ and consider the regime $0<\epsilon\ll 1$.  Rewriting \eqref{eq1} in terms of the small parameter $\epsilon$ and performing a formal multiple-scale expansion in the weakly nonlinear regime yields a generalized \textit{focusing} nonlinear Schr\"odinger (NLS) equation for the breather's spatial envelope when the nonlinearity is \textit{softening} (see Appendix B). Such an asymptotic analysis yields the following expressions for small-amplitude breather norms:
\begin{align} 
\label{norms}
&\|\{x_{n,m}^{J}(0)\}_{\{(n,m)\in\mathbb{Z}^2,J\}}\|_{\ell^{\infty}}\sim 2\epsilon^{1/2\sigma}\|S_0\|_{L^{\infty}(\mathbb{R}^2)}
\\\nonumber
&\|\{x_{n,m}^{J}(0)\}_{\{(n,m)\in\mathbb{Z}^2,J\}}\|_{\ell^2}^2\sim 2\sqrt{3}\epsilon^{(1-\sigma)/\sigma}\|S_0\|_{L^{2}(\mathbb{R}^2)}^2,
\end{align}
\newline
\noindent for $J\in(A,B,C)$ and where $S_0(Z,H)$ is the radial ground-state solitary wave solution of the derived NLS equation and can be obtained numerically~\cite{fibich}.  Our asymptotic expressions for the norms are shown by the solid curves in Figure \ref{figure4} for two different strengths of nonlinearity $(\sigma=1,2).$  The dots in Figure \ref{figure4} are again the corresponding breather norms, numerically continued from the distant molecular limit. As expected, the dots and solid curves begin to merge near the band edge as $\lambda$ approaches $\lambda_*$.  In the case of cubic nonlinearity, $\sigma=1$, the $\ell^2$-scaling is \textit{critical}, meaning that breathers bifurcating from the zero solution exist only above a certain nonzero finite energy threshold, which is indicated with the solid horizontal line in Figure \ref{figure4}(a).  This energy threshold becomes infinite when considering the \textit{super-critical} case, $\sigma=2$, shown in Figure \ref{figure4}(b). In this case, the lattice does not support small-amplitude discrete breathers arbitrarily close to the phonon band.  We are only able to compute breathers within tolerance up to the point shown nearest $\lambda_*$.  We expect our asymptotic analysis to improve as $\lambda\to\lambda_*$.
\subsubsection{Breathers Above the Optical Bands}
Finally, as a result of the flat phonon band~\cite{flach2}, the BK lattice admits discrete breathers that identically vanish
\onecolumngrid
\begin{figure}[t!]\setlength{\hfuzz}{1.1\columnwidth}
\begin{minipage}{\textwidth}
\ttfamily
    \includegraphics[width=16cm,height=16cm,keepaspectratio,angle=90]{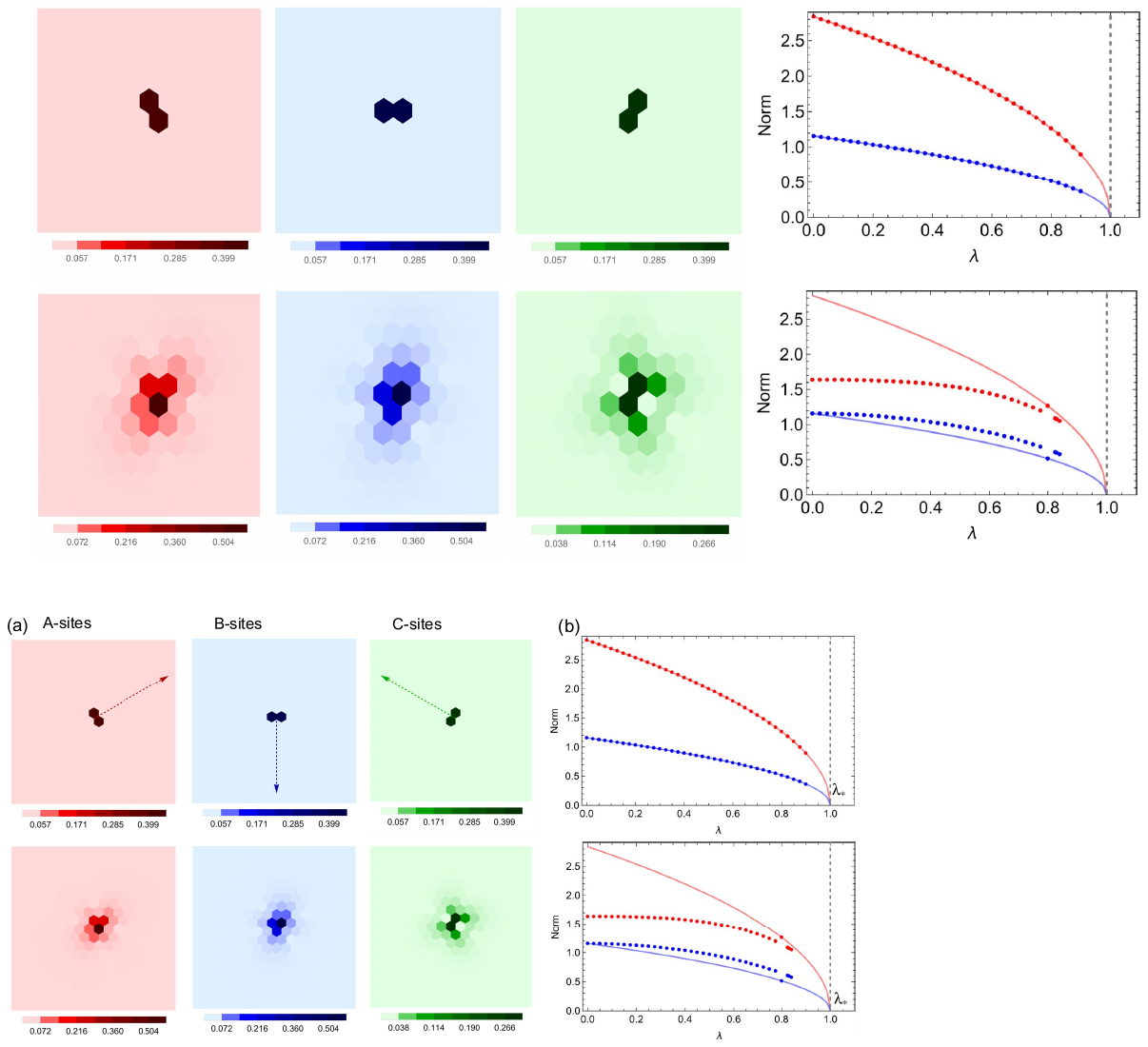}
    \caption{\textbf{Top row of panels}: (S) six-site compact breather in \eqref{hex}-\eqref{ellip} with $\omega_b$ above flat phonon band (a) absolute spatial profile when $\lambda=0.9$ (arrows highlight $C_3$ sub-lattice symmetry); (b) infinity- (blue dots) and 2- (red dots) norms computed from molecular limit discussed in the text and solid curves are the exact norms ($\lambda_*=1$ dashed vertical line) \textbf{Bottom row of panels}: (A) same as top row but seeded from a \textit{single} out-of-phase state, where $\lambda=0.84$ in (a) and the solid curves in the bottom panel of (b) are copied from the top panel for reference. (parameters: $g=1$, $\omega_b=6$).}
    \label{figure5}
    \end{minipage}
\end{figure}
\twocolumngrid
\noindent off of a finite subset of lattice sites, i.e., having compact support. An explicit solution of this type is supported on the six lattice sites of a closed hexagonal ring, given by
\begin{align}\label{hex}
&x_{0,0}^A=x_{-1,1}^C=x_{-1,0}^B=z^{(II)}(t),\\ \nonumber
&x_{0,0}^B=x_{-1,1}^A=x_{-1,0}^C=-z^{(II)}(t),\\ \nonumber
&x_{n,m}^J\equiv 0\text{ for all other $(n,m)\in\mathbb{Z}^2$,}
\end{align}
where $z^{(II)}(t)$ is a $T_b$-periodic solution to the initial value problem \eqref{out}, with $\gamma$ replaced by $\gamma+\lambda$. In the case of a cubic nonlinearity, \eqref{out} can be solved exactly in terms of the even Jacobi elliptic function:
\begin{equation}\label{ellip1}
    z^{(II)}(t)=a_*\text{cn}\left(\sqrt{(s+ga_*^2)}t,\,\dfrac{ga_*^2}{2(s+ga_*^2)}\right)
\end{equation}
with a period given by
\begin{equation}\label{ellip}
T_b=\dfrac{4}{\sqrt{s+ga_*^2}}\mathcal{K}\left(\sqrt{\dfrac{ga_*^2}{2\left(s+ga_*^2\right)}}\right),
\end{equation} 
where $s:=\omega_0^2+\lambda+\gamma$ and $\mathcal{K}$ is the complete elliptic integral of the first kind~\cite{stegun}. Similar compact breathers were obtained in a dNLS model of a Kagome lattice in~\cite{johansson}.   

In Figure \ref{figure5} we consider breather families that have a fixed frequency above the flat phonon band and $g>0$.  The upper panels of Figure \ref{figure5} show breathers given by equations \eqref{hex}-\eqref{ellip}. A representative example of the compact spatial profile of the absolute values of this breather is depicted in the upper panels of Figure \ref{figure5}(a) (we refer to these breathers by (S), for symmetric).  Here again, the three plots show the projections of the breather onto each of its sub-lattices. The solid curves in Figure \ref{figure5}(b) show the exact norms of (S) breathers versus $\lambda$, while the overlaying points validate our numerical implementation.  In this case, there are breathers of arbitrarily small energy, unlike those, for example, shown in Figure \ref{figure4}.  We remark that, at the molecular limit ($\lambda=0$), (S) breathers consist of \textit{three} out-of-phase states arranged along a hexagonal ring. 

\begin{figure}[t!]
    \includegraphics[scale=0.61]{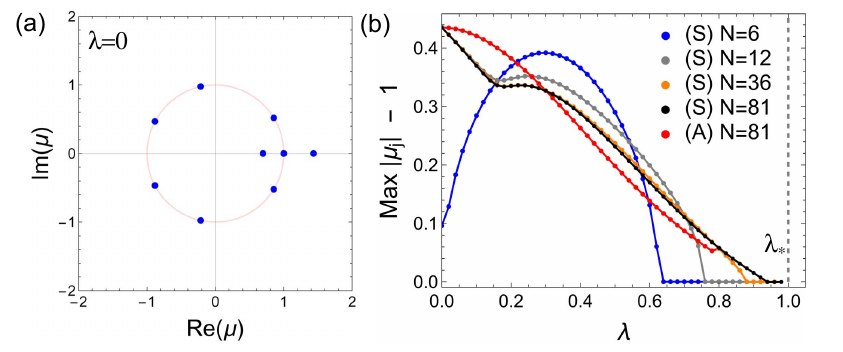}
    \caption{(a) Floquet spectrum of the (S) and (A) breathers in Figure \ref{figure5} at $\lambda=0$; (b) maximum deviation of the Floquet spectrum's modulus from one versus $\lambda$ of (S) and (A) breathers for different lattice sizes $N$.}
    \label{figure6}
\end{figure}

In the bottom panels of Figure \ref{figure5}, the numerical continuation scheme is seeded with a \textit{single} out-of-phase state in the molecular limit (we refer to these breathers as (A), for asymmetric). The family of (A) breathers lacks the rotational symmetry of the (S) family but remains extremely localized in the globally coupled lattice and appears to maintain compact support.  As the points in Figure \ref{figure5}(b) show, the (A) breathers also appear to lack an energy threshold as $\lambda\to\lambda_*$.  We note that at $\lambda\approx 0.8$ in Figure \ref{figure5}(b), the (A) continuation branch briefly switches to the (S) branch.  Both branches are dynamically unstable over this region for the given parameters and lattice size (see Figure \ref{figure6}).  Our numerical iteration scheme fails to converge with a finer $\lambda$- resolution just before and after this transition.

Figure \ref{figure6} indicates the dynamic stability of the two breather families of Figure \ref{figure5}.  At the molecular limit, both families are unstable and possess a pair of Floquet exponents outside the unit circle on the real-axis as shown in Figure \ref{figure6}(a). Figure \ref{figure6}(b) shows that, as $\lambda$ approaches $\lambda_*$, the exact (S) family has a window of stability.  The stability region near the edge of the flat band in the weakly nonlinear regime is similar to the one reported in~\cite{johansson} for breathers in a dNLS model. The width of the stability region depends on the size and boundary of the truncated lattice.  Here, we impose zero Dirichlet boundary conditions.
Figure \ref{figure6}(b) shows that as the number of lattice sites increases around the central (S) breather in \eqref{hex}, the stability window shrinks.  The decreasing stability window coincides with an increase in the number of standing waves the lattice supports. Figure \ref{figure6}(b) shows that the (A) family is unstable and its numerical continuation merges with the (S) family near $\lambda=0.8.$  Addressing the stability switching, localization, and dynamic properties of these compact nonlinear modes near the flat band more deeply is an open avenue for further research. 

\section{Conclusion}
In this article, we have shown that the unique geometry of the Kagome lattice offers a novel means to shape diverse families of discrete breathers. Recently, there has been a tremendous amount of interest in higher-order topological properties inherent in the linear BK lattice \cite{khanikaev,ezawa,berc}, leading to the formation of protected edge-modes and corner states along a truncated lattice.  Here, we have focused on families of intrinsically localized modes inside the bulk of the lattice.  An interesting extension of the work presented here would be determining whether there exists a connection between discrete breathers, seeded from the molecular limit, and nonlinear continuations of corner states on the lattice with truncated boundaries~\cite{theo}. Another interesting direction is the broader study of coherent structures in novel temporal and spatiotemporal engineered crystal lattices that exhibit periodicity in time and space-time, respectively. Such time-dependent materials are capable of opening band gaps in both energy and momentum space~\cite{sza}.   In summary, the results presented here provide a means to precisely control energy localization and stabilization of coherent structures in the bulk of the BK lattice via nonlinearity and easily tunable parameters in physical systems.

\section*{Acknowledgments}
This work was supported by the Air Force Office of Scientific Research with Grant No. FA9550-23-1-0144 and the Simons Foundation collaboration ``Harnessing Universal Symmetry Concepts for Extreme Wave Phenomena."
\bibliography{apssamp}

\providecommand{\noopsort}[1]{}\providecommand{\singleletter}[1]{#1}%
\begin{thebibliography}{41}%
\makeatletter
\providecommand \@ifxundefined [1]{%
 \@ifx{#1\undefined}
}%
\providecommand \@ifnum [1]{%
 \ifnum #1\expandafter \@firstoftwo
 \else \expandafter \@secondoftwo
 \fi
}%
\providecommand \@ifx [1]{%
 \ifx #1\expandafter \@firstoftwo
 \else \expandafter \@secondoftwo
 \fi
}%
\providecommand \natexlab [1]{#1}%
\providecommand \enquote  [1]{``#1''}%
\providecommand \bibnamefont  [1]{#1}%
\providecommand \bibfnamefont [1]{#1}%
\providecommand \citenamefont [1]{#1}%
\providecommand \href@noop [0]{\@secondoftwo}%
\providecommand \href [0]{\begingroup \@sanitize@url \@href}%
\providecommand \@href[1]{\@@startlink{#1}\@@href}%
\providecommand \@@href[1]{\endgroup#1\@@endlink}%
\providecommand \@sanitize@url [0]{\catcode `\\12\catcode `\$12\catcode `\&12\catcode `\#12\catcode `\^12\catcode `\_12\catcode `\%12\relax}%
\providecommand \@@startlink[1]{}%
\providecommand \@@endlink[0]{}%
\providecommand \url  [0]{\begingroup\@sanitize@url \@url }%
\providecommand \@url [1]{\endgroup\@href {#1}{\urlprefix }}%
\providecommand \urlprefix  [0]{URL }%
\providecommand \Eprint [0]{\href }%
\providecommand \doibase [0]{https://doi.org/}%
\providecommand \selectlanguage [0]{\@gobble}%
\providecommand \bibinfo  [0]{\@secondoftwo}%
\providecommand \bibfield  [0]{\@secondoftwo}%
\providecommand \translation [1]{[#1]}%
\providecommand \BibitemOpen [0]{}%
\providecommand \bibitemStop [0]{}%
\providecommand \bibitemNoStop [0]{.\EOS\space}%
\providecommand \EOS [0]{\spacefactor3000\relax}%
\providecommand \BibitemShut  [1]{\csname bibitem#1\endcsname}%
\let\auto@bib@innerbib\@empty
\bibitem [{\citenamefont {Braun}\ and\ \citenamefont {Kivshar}(2010)}]{kivshar}%
  \BibitemOpen
  \bibfield  {author} {\bibinfo {author} {\bibfnamefont {O.~M.}\ \bibnamefont {Braun}}\ and\ \bibinfo {author} {\bibfnamefont {Y.~S.}\ \bibnamefont {Kivshar}},\ }\href@noop {} {\emph {\bibinfo {title} {Frenkel Kontorova Model: Concepts, Methods, and Applications}}}\ (\bibinfo  {publisher} {Springer},\ \bibinfo {year} {2010})\BibitemShut {NoStop}%
\bibitem [{\citenamefont {Morsch}\ and\ \citenamefont {Oberthaler}(2006)}]{ober}%
  \BibitemOpen
  \bibfield  {author} {\bibinfo {author} {\bibfnamefont {O.}~\bibnamefont {Morsch}}\ and\ \bibinfo {author} {\bibfnamefont {M.}~\bibnamefont {Oberthaler}},\ }\bibfield  {title} {\bibinfo {title} {Dynamics of bose-einstein condensates in optical lattices},\ }\href@noop {} {\bibfield  {journal} {\bibinfo  {journal} {Rev. Mod. Phys.}\ }\textbf {\bibinfo {volume} {78}} (\bibinfo {year} {2006})}\BibitemShut {NoStop}%
\bibitem [{\citenamefont {Moloney}\ and\ \citenamefont {Newell}(2003)}]{moloney}%
  \BibitemOpen
  \bibfield  {author} {\bibinfo {author} {\bibfnamefont {J.~V.}\ \bibnamefont {Moloney}}\ and\ \bibinfo {author} {\bibfnamefont {A.~C.}\ \bibnamefont {Newell}},\ }\href@noop {} {\emph {\bibinfo {title} {Nonlinear Optics}}}\ (\bibinfo  {publisher} {CRC Press LLC},\ \bibinfo {year} {2003})\BibitemShut {NoStop}%
\bibitem [{\citenamefont {Blanco-Redondo}\ \emph {et~al.}(2023)\citenamefont {Blanco-Redondo}, \citenamefont {de~Sterke}, \citenamefont {Xu}, \citenamefont {Wabnitz},\ and\ \citenamefont {Turitsyn}}]{turitsyn}%
  \BibitemOpen
  \bibfield  {author} {\bibinfo {author} {\bibfnamefont {A.}~\bibnamefont {Blanco-Redondo}}, \bibinfo {author} {\bibfnamefont {C.~M.}\ \bibnamefont {de~Sterke}}, \bibinfo {author} {\bibfnamefont {C.}~\bibnamefont {Xu}}, \bibinfo {author} {\bibfnamefont {S.}~\bibnamefont {Wabnitz}},\ and\ \bibinfo {author} {\bibfnamefont {S.~K.}\ \bibnamefont {Turitsyn}},\ }\bibfield  {title} {\bibinfo {title} {The bright prospects of optical solitons after 50 years},\ }\href@noop {} {\bibfield  {journal} {\bibinfo  {journal} {Nature Photonics}\ }\textbf {\bibinfo {volume} {17}} (\bibinfo {year} {2023})}\BibitemShut {NoStop}%
\bibitem [{\citenamefont {Bertoldi}\ \emph {et~al.}(2017)\citenamefont {Bertoldi}, \citenamefont {Vitelli}, \citenamefont {Christensen},\ and\ \citenamefont {van Hecke}}]{hecke}%
  \BibitemOpen
  \bibfield  {author} {\bibinfo {author} {\bibfnamefont {K.}~\bibnamefont {Bertoldi}}, \bibinfo {author} {\bibfnamefont {V.}~\bibnamefont {Vitelli}}, \bibinfo {author} {\bibfnamefont {J.}~\bibnamefont {Christensen}},\ and\ \bibinfo {author} {\bibfnamefont {M.}~\bibnamefont {van Hecke}},\ }\bibfield  {title} {\bibinfo {title} {Flexible mechanical metamaterials},\ }\href@noop {} {\bibfield  {journal} {\bibinfo  {journal} {Nature Reviews Materials}\ }\textbf {\bibinfo {volume} {2}} (\bibinfo {year} {2017})}\BibitemShut {NoStop}%
\bibitem [{\citenamefont {Zhang}\ \emph {et~al.}(2019)\citenamefont {Zhang}, \citenamefont {Li}, \citenamefont {Zheng}, \citenamefont {Genin},\ and\ \citenamefont {Chen}}]{chen}%
  \BibitemOpen
  \bibfield  {author} {\bibinfo {author} {\bibfnamefont {Y.}~\bibnamefont {Zhang}}, \bibinfo {author} {\bibfnamefont {B.}~\bibnamefont {Li}}, \bibinfo {author} {\bibfnamefont {Q.~S.}\ \bibnamefont {Zheng}}, \bibinfo {author} {\bibfnamefont {G.~M.}\ \bibnamefont {Genin}},\ and\ \bibinfo {author} {\bibfnamefont {C.~Q.}\ \bibnamefont {Chen}},\ }\bibfield  {title} {\bibinfo {title} {Programmable and robust static topological solitons in mechanical metamaterials},\ }\href@noop {} {\bibfield  {journal} {\bibinfo  {journal} {Nature Communications}\ }\textbf {\bibinfo {volume} {10}} (\bibinfo {year} {2019})}\BibitemShut {NoStop}%
\bibitem [{\citenamefont {Hadad}\ \emph {et~al.}(2018)\citenamefont {Hadad}, \citenamefont {Soric}, \citenamefont {Khanikaev},\ and\ \citenamefont {Alu}}]{alu}%
  \BibitemOpen
  \bibfield  {author} {\bibinfo {author} {\bibfnamefont {Y.}~\bibnamefont {Hadad}}, \bibinfo {author} {\bibfnamefont {J.~C.}\ \bibnamefont {Soric}}, \bibinfo {author} {\bibfnamefont {A.~B.}\ \bibnamefont {Khanikaev}},\ and\ \bibinfo {author} {\bibfnamefont {A.}~\bibnamefont {Alu}},\ }\bibfield  {title} {\bibinfo {title} {Self-induced topological protection in nonlinear circuit arrays},\ }\href@noop {} {\bibfield  {journal} {\bibinfo  {journal} {Nature Electronics}\ }\textbf {\bibinfo {volume} {1}} (\bibinfo {year} {2018})}\BibitemShut {NoStop}%
\bibitem [{\citenamefont {Li}\ \emph {et~al.}(2025)\citenamefont {Li}, \citenamefont {Hofstrand},\ and\ \citenamefont {Weinstein}}]{li}%
  \BibitemOpen
  \bibfield  {author} {\bibinfo {author} {\bibfnamefont {H.}~\bibnamefont {Li}}, \bibinfo {author} {\bibfnamefont {A.}~\bibnamefont {Hofstrand}},\ and\ \bibinfo {author} {\bibfnamefont {M.~I.}\ \bibnamefont {Weinstein}},\ }\bibfield  {title} {\bibinfo {title} {Stability of traveling waves in a nonlinear hyperbolic system approximating a dimer array of oscillators},\ }\href@noop {} {\bibfield  {journal} {\bibinfo  {journal} {Journal of Nonlinear Science}\ }\textbf {\bibinfo {volume} {35}} (\bibinfo {year} {2025})}\BibitemShut {NoStop}%
\bibitem [{\citenamefont {MacKay}\ and\ \citenamefont {Aubry}(1994)}]{aubry}%
  \BibitemOpen
  \bibfield  {author} {\bibinfo {author} {\bibfnamefont {R.~S.}\ \bibnamefont {MacKay}}\ and\ \bibinfo {author} {\bibfnamefont {S.}~\bibnamefont {Aubry}},\ }\bibfield  {title} {\bibinfo {title} {Proof of existence of breathers for time-reversible or hamiltonian networks of weakly coupled oscillators.},\ }\href@noop {} {\bibfield  {journal} {\bibinfo  {journal} {Nonlinearity}\ }\textbf {\bibinfo {volume} {7}} (\bibinfo {year} {1994})}\BibitemShut {NoStop}%
\bibitem [{\citenamefont {Zabusky}\ \emph {et~al.}(2006)\citenamefont {Zabusky}, \citenamefont {Sun},\ and\ \citenamefont {Peng}}]{zabusky}%
  \BibitemOpen
  \bibfield  {author} {\bibinfo {author} {\bibfnamefont {N.~J.}\ \bibnamefont {Zabusky}}, \bibinfo {author} {\bibfnamefont {Z.}~\bibnamefont {Sun}},\ and\ \bibinfo {author} {\bibfnamefont {G.}~\bibnamefont {Peng}},\ }\bibfield  {title} {\bibinfo {title} {Measures of chaos and equipartition in integrable and nonintegrable lattices},\ }\href@noop {} {\bibfield  {journal} {\bibinfo  {journal} {Chaos}\ }\textbf {\bibinfo {volume} {16}} (\bibinfo {year} {2006})}\BibitemShut {NoStop}%
\bibitem [{\citenamefont {Xiong}\ \emph {et~al.}(2012)\citenamefont {Xiong}, \citenamefont {Wang}, \citenamefont {Zhang},\ and\ \citenamefont {Zhao}}]{xiong1}%
  \BibitemOpen
  \bibfield  {author} {\bibinfo {author} {\bibfnamefont {D.}~\bibnamefont {Xiong}}, \bibinfo {author} {\bibfnamefont {J.}~\bibnamefont {Wang}}, \bibinfo {author} {\bibfnamefont {Y.}~\bibnamefont {Zhang}},\ and\ \bibinfo {author} {\bibfnamefont {H.}~\bibnamefont {Zhao}},\ }\bibfield  {title} {\bibinfo {title} {Nonuniversal heat conduction of one-dimensional lattices},\ }\href@noop {} {\bibfield  {journal} {\bibinfo  {journal} {Physical Review E}\ }\textbf {\bibinfo {volume} {85}} (\bibinfo {year} {2012})}\BibitemShut {NoStop}%
\bibitem [{\citenamefont {Xiong}\ \emph {et~al.}(2017)\citenamefont {Xiong}, \citenamefont {Saadatmand},\ and\ \citenamefont {Dmitriev}}]{xiong2}%
  \BibitemOpen
  \bibfield  {author} {\bibinfo {author} {\bibfnamefont {D.}~\bibnamefont {Xiong}}, \bibinfo {author} {\bibfnamefont {D.}~\bibnamefont {Saadatmand}},\ and\ \bibinfo {author} {\bibfnamefont {S.~V.}\ \bibnamefont {Dmitriev}},\ }\bibfield  {title} {\bibinfo {title} {Crossover from ballistic to normal heat transport in the $\phi^4$ lattice},\ }\href@noop {} {\bibfield  {journal} {\bibinfo  {journal} {Physical Review E}\ }\textbf {\bibinfo {volume} {96}} (\bibinfo {year} {2017})}\BibitemShut {NoStop}%
\bibitem [{\citenamefont {Caballero}\ and\ \citenamefont {Wayne}(2022)}]{wayne}%
  \BibitemOpen
  \bibfield  {author} {\bibinfo {author} {\bibfnamefont {D.~A.}\ \bibnamefont {Caballero}}\ and\ \bibinfo {author} {\bibfnamefont {C.~E.}\ \bibnamefont {Wayne}},\ }\bibfield  {title} {\bibinfo {title} {Damped and driven breathers and metastability},\ }\bibfield  {journal} {\bibinfo  {journal} {arXiv}\ }\href {https://doi.org/https://doi.org/10.48550/arXiv.2101.10999} {https://doi.org/10.48550/arXiv.2101.10999} (\bibinfo {year} {2022})\BibitemShut {NoStop}%
\bibitem [{\citenamefont {Flach}\ and\ \citenamefont {Gorbach}(2008)}]{flach}%
  \BibitemOpen
  \bibfield  {author} {\bibinfo {author} {\bibfnamefont {S.}~\bibnamefont {Flach}}\ and\ \bibinfo {author} {\bibfnamefont {A.~V.}\ \bibnamefont {Gorbach}},\ }\bibfield  {title} {\bibinfo {title} {Discrete breathers—advances in theory and applications},\ }\href@noop {} {\bibfield  {journal} {\bibinfo  {journal} {Phys. Rep.}\ }\textbf {\bibinfo {volume} {467}} (\bibinfo {year} {2008})}\BibitemShut {NoStop}%
\bibitem [{\citenamefont {Lustri}\ and\ \citenamefont {Porter}(2018)}]{porter}%
  \BibitemOpen
  \bibfield  {author} {\bibinfo {author} {\bibfnamefont {C.~J.}\ \bibnamefont {Lustri}}\ and\ \bibinfo {author} {\bibfnamefont {M.~A.}\ \bibnamefont {Porter}},\ }\bibfield  {title} {\bibinfo {title} {Nanoptera in a period-2 toda chain},\ }\href@noop {} {\bibfield  {journal} {\bibinfo  {journal} {SIAM Journal on Applied Dynamical Systems}\ }\textbf {\bibinfo {volume} {17}} (\bibinfo {year} {2018})}\BibitemShut {NoStop}%
\bibitem [{\citenamefont {Liu}\ \emph {et~al.}(2024)\citenamefont {Liu}, \citenamefont {Liu}, \citenamefont {Zhang}, \citenamefont {Sun}, \citenamefont {Zhang}, \citenamefont {Wang},\ and\ \citenamefont {Liu}}]{liu}%
  \BibitemOpen
  \bibfield  {author} {\bibinfo {author} {\bibfnamefont {C.}~\bibnamefont {Liu}}, \bibinfo {author} {\bibfnamefont {T.}~\bibnamefont {Liu}}, \bibinfo {author} {\bibfnamefont {Z.}~\bibnamefont {Zhang}}, \bibinfo {author} {\bibfnamefont {Z.}~\bibnamefont {Sun}}, \bibinfo {author} {\bibfnamefont {G.}~\bibnamefont {Zhang}}, \bibinfo {author} {\bibfnamefont {E.}~\bibnamefont {Wang}},\ and\ \bibinfo {author} {\bibfnamefont {K.}~\bibnamefont {Liu}},\ }\bibfield  {title} {\bibinfo {title} {Understanding epitaxial growth of two-dimensional materials and their homostructures},\ }\href@noop {} {\bibfield  {journal} {\bibinfo  {journal} {Nature Nanotechnology}\ } (\bibinfo {year} {2024})}\BibitemShut {NoStop}%
\bibitem [{\citenamefont {Jo}\ \emph {et~al.}(2012)\citenamefont {Jo}, \citenamefont {Guzman}, \citenamefont {Thomas}, \citenamefont {Hosur}, \citenamefont {Vishwanath},\ and\ \citenamefont {Stamper-Kurn}}]{kurn0}%
  \BibitemOpen
  \bibfield  {author} {\bibinfo {author} {\bibfnamefont {G.-B.}\ \bibnamefont {Jo}}, \bibinfo {author} {\bibfnamefont {J.}~\bibnamefont {Guzman}}, \bibinfo {author} {\bibfnamefont {C.~K.}\ \bibnamefont {Thomas}}, \bibinfo {author} {\bibfnamefont {P.}~\bibnamefont {Hosur}}, \bibinfo {author} {\bibfnamefont {A.}~\bibnamefont {Vishwanath}},\ and\ \bibinfo {author} {\bibfnamefont {D.~M.}\ \bibnamefont {Stamper-Kurn}},\ }\bibfield  {title} {\bibinfo {title} {Ultracold atoms in a tunable optical kagome lattice},\ }\href@noop {} {\bibfield  {journal} {\bibinfo  {journal} {Physical Review Letters}\ }\textbf {\bibinfo {volume} {108}} (\bibinfo {year} {2012})}\BibitemShut {NoStop}%
\bibitem [{\citenamefont {Yin}\ \emph {et~al.}(2022)\citenamefont {Yin}, \citenamefont {Lian},\ and\ \citenamefont {Hasan}}]{hasan}%
  \BibitemOpen
  \bibfield  {author} {\bibinfo {author} {\bibfnamefont {J.-X.}\ \bibnamefont {Yin}}, \bibinfo {author} {\bibfnamefont {B.}~\bibnamefont {Lian}},\ and\ \bibinfo {author} {\bibfnamefont {M.~Z.}\ \bibnamefont {Hasan}},\ }\bibfield  {title} {\bibinfo {title} {Topological kagome magnets and superconductors},\ }\href@noop {} {\bibfield  {journal} {\bibinfo  {journal} {Nature}\ }\textbf {\bibinfo {volume} {612}} (\bibinfo {year} {2022})}\BibitemShut {NoStop}%
\bibitem [{\citenamefont {Fruchart}\ \emph {et~al.}(2020)\citenamefont {Fruchart}, \citenamefont {Zhou},\ and\ \citenamefont {Vitelli}}]{vitelli}%
  \BibitemOpen
  \bibfield  {author} {\bibinfo {author} {\bibfnamefont {M.}~\bibnamefont {Fruchart}}, \bibinfo {author} {\bibfnamefont {Y.}~\bibnamefont {Zhou}},\ and\ \bibinfo {author} {\bibfnamefont {V.}~\bibnamefont {Vitelli}},\ }\bibfield  {title} {\bibinfo {title} {Dualities and non-abelian mechanics},\ }\href@noop {} {\bibfield  {journal} {\bibinfo  {journal} {Nature}\ }\textbf {\bibinfo {volume} {577}} (\bibinfo {year} {2020})}\BibitemShut {NoStop}%
\bibitem [{\citenamefont {Li}\ \emph {et~al.}(2020)\citenamefont {Li}, \citenamefont {Zhirihin}, \citenamefont {Gorlach}, \citenamefont {Ni}, \citenamefont {Filonov}, \citenamefont {Slobozhanyuk}, \citenamefont {Alu},\ and\ \citenamefont {Khanikaev}}]{khan}%
  \BibitemOpen
  \bibfield  {author} {\bibinfo {author} {\bibfnamefont {M.}~\bibnamefont {Li}}, \bibinfo {author} {\bibfnamefont {D.}~\bibnamefont {Zhirihin}}, \bibinfo {author} {\bibfnamefont {M.}~\bibnamefont {Gorlach}}, \bibinfo {author} {\bibfnamefont {X.}~\bibnamefont {Ni}}, \bibinfo {author} {\bibfnamefont {D.}~\bibnamefont {Filonov}}, \bibinfo {author} {\bibfnamefont {A.}~\bibnamefont {Slobozhanyuk}}, \bibinfo {author} {\bibfnamefont {A.}~\bibnamefont {Alu}},\ and\ \bibinfo {author} {\bibfnamefont {A.~B.}\ \bibnamefont {Khanikaev}},\ }\bibfield  {title} {\bibinfo {title} {Higher-order topological states in photonic kagome crystals with long-range interactions},\ }\href@noop {} {\bibfield  {journal} {\bibinfo  {journal} {Nature Photonics}\ }\textbf {\bibinfo {volume} {14}} (\bibinfo {year} {2020})}\BibitemShut {NoStop}%
\bibitem [{\citenamefont {Hofhuis}\ \emph {et~al.}(2022)\citenamefont {Hofhuis}, \citenamefont {Skjaervo}, \citenamefont {Parchenko}, \citenamefont {Arava}, \citenamefont {Luo}, \citenamefont {Kleibert}, \citenamefont {Derlet},\ and\ \citenamefont {Heyderman}}]{heyderman}%
  \BibitemOpen
  \bibfield  {author} {\bibinfo {author} {\bibfnamefont {K.}~\bibnamefont {Hofhuis}}, \bibinfo {author} {\bibfnamefont {S.~H.}\ \bibnamefont {Skjaervo}}, \bibinfo {author} {\bibfnamefont {S.}~\bibnamefont {Parchenko}}, \bibinfo {author} {\bibfnamefont {H.}~\bibnamefont {Arava}}, \bibinfo {author} {\bibfnamefont {Z.}~\bibnamefont {Luo}}, \bibinfo {author} {\bibfnamefont {A.}~\bibnamefont {Kleibert}}, \bibinfo {author} {\bibfnamefont {P.~M.}\ \bibnamefont {Derlet}},\ and\ \bibinfo {author} {\bibfnamefont {L.~J.}\ \bibnamefont {Heyderman}},\ }\bibfield  {title} {\bibinfo {title} {Real-space imaging of phase transitions in bridged artificial kagome spin ice},\ }\href@noop {} {\bibfield  {journal} {\bibinfo  {journal} {Nature Physics}\ }\textbf {\bibinfo {volume} {18}} (\bibinfo {year} {2022})}\BibitemShut {NoStop}%
\bibitem [{\citenamefont {Bergman}\ \emph {et~al.}(2008)\citenamefont {Bergman}, \citenamefont {Wu},\ and\ \citenamefont {Balents}}]{balents}%
  \BibitemOpen
  \bibfield  {author} {\bibinfo {author} {\bibfnamefont {D.~L.}\ \bibnamefont {Bergman}}, \bibinfo {author} {\bibfnamefont {C.}~\bibnamefont {Wu}},\ and\ \bibinfo {author} {\bibfnamefont {L.}~\bibnamefont {Balents}},\ }\bibfield  {title} {\bibinfo {title} {Band touching from real-space topology in frustrated hopping models},\ }\href@noop {} {\bibfield  {journal} {\bibinfo  {journal} {Physical Review B}\ }\textbf {\bibinfo {volume} {78}} (\bibinfo {year} {2008})}\BibitemShut {NoStop}%
\bibitem [{\citenamefont {Ni}\ \emph {et~al.}(2019)\citenamefont {Ni}, \citenamefont {Weiner}, \citenamefont {Alu},\ and\ \citenamefont {Khanikaev}}]{khanikaev}%
  \BibitemOpen
  \bibfield  {author} {\bibinfo {author} {\bibfnamefont {X.}~\bibnamefont {Ni}}, \bibinfo {author} {\bibfnamefont {M.}~\bibnamefont {Weiner}}, \bibinfo {author} {\bibfnamefont {A.}~\bibnamefont {Alu}},\ and\ \bibinfo {author} {\bibfnamefont {A.~B.}\ \bibnamefont {Khanikaev}},\ }\bibfield  {title} {\bibinfo {title} {Observation of higher-order topological acoustic states protected by generalized chiral symmetry},\ }\href@noop {} {\bibfield  {journal} {\bibinfo  {journal} {Nature Materials}\ }\textbf {\bibinfo {volume} {18}} (\bibinfo {year} {2019})}\BibitemShut {NoStop}%
\bibitem [{\citenamefont {Herrera}\ \emph {et~al.}(2022)\citenamefont {Herrera}, \citenamefont {Kempkes}, \citenamefont {de~Paz}, \citenamefont {García-Etxarri}, \citenamefont {Swart}, \citenamefont {Smith},\ and\ \citenamefont {Bercioux}}]{berc}%
  \BibitemOpen
  \bibfield  {author} {\bibinfo {author} {\bibfnamefont {M.~A.~J.}\ \bibnamefont {Herrera}}, \bibinfo {author} {\bibfnamefont {S.~N.}\ \bibnamefont {Kempkes}}, \bibinfo {author} {\bibfnamefont {M.~B.}\ \bibnamefont {de~Paz}}, \bibinfo {author} {\bibfnamefont {A.}~\bibnamefont {García-Etxarri}}, \bibinfo {author} {\bibfnamefont {I.}~\bibnamefont {Swart}}, \bibinfo {author} {\bibfnamefont {C.~M.}\ \bibnamefont {Smith}},\ and\ \bibinfo {author} {\bibfnamefont {D.}~\bibnamefont {Bercioux}},\ }\bibfield  {title} {\bibinfo {title} {Corner modes of the breathing kagome lattice: Origin and robustness},\ }\href@noop {} {\bibfield  {journal} {\bibinfo  {journal} {Physical Review B}\ }\textbf {\bibinfo {volume} {105}} (\bibinfo {year} {2022})}\BibitemShut {NoStop}%
\bibitem [{\citenamefont {Wang}\ \emph {et~al.}(2023)\citenamefont {Wang}, \citenamefont {Wang}, \citenamefont {Hu}, \citenamefont {Bongiovanni}, \citenamefont {Jukić}, \citenamefont {Tang}, \citenamefont {Song}, \citenamefont {Morandotti}, \citenamefont {Chen},\ and\ \citenamefont {Buljan}}]{buljan}%
  \BibitemOpen
  \bibfield  {author} {\bibinfo {author} {\bibfnamefont {Z.}~\bibnamefont {Wang}}, \bibinfo {author} {\bibfnamefont {X.}~\bibnamefont {Wang}}, \bibinfo {author} {\bibfnamefont {Z.}~\bibnamefont {Hu}}, \bibinfo {author} {\bibfnamefont {D.}~\bibnamefont {Bongiovanni}}, \bibinfo {author} {\bibfnamefont {D.}~\bibnamefont {Jukić}}, \bibinfo {author} {\bibfnamefont {L.}~\bibnamefont {Tang}}, \bibinfo {author} {\bibfnamefont {D.}~\bibnamefont {Song}}, \bibinfo {author} {\bibfnamefont {R.}~\bibnamefont {Morandotti}}, \bibinfo {author} {\bibfnamefont {Z.}~\bibnamefont {Chen}},\ and\ \bibinfo {author} {\bibfnamefont {H.}~\bibnamefont {Buljan}},\ }\bibfield  {title} {\bibinfo {title} {Sub-symmetry-protected topological states},\ }\href@noop {} {\bibfield  {journal} {\bibinfo  {journal} {Nature Physics}\ }\textbf {\bibinfo {volume} {19}} (\bibinfo {year} {2023})}\BibitemShut {NoStop}%
\bibitem [{\citenamefont {Ezawa}(2018)}]{ezawa}%
  \BibitemOpen
  \bibfield  {author} {\bibinfo {author} {\bibfnamefont {M.}~\bibnamefont {Ezawa}},\ }\bibfield  {title} {\bibinfo {title} {Higher-order topological insulators and semimetals on the breathing kagome and pyrochlore lattices},\ }\href@noop {} {\bibfield  {journal} {\bibinfo  {journal} {Physical Review Letters}\ }\textbf {\bibinfo {volume} {120}} (\bibinfo {year} {2018})}\BibitemShut {NoStop}%
\bibitem [{\citenamefont {Barter}\ \emph {et~al.}(2020)\citenamefont {Barter}, \citenamefont {Leung}, \citenamefont {Okano}, \citenamefont {Block}, \citenamefont {Yao},\ and\ \citenamefont {Stamper-Kurn}}]{kurn}%
  \BibitemOpen
  \bibfield  {author} {\bibinfo {author} {\bibfnamefont {T.~H.}\ \bibnamefont {Barter}}, \bibinfo {author} {\bibfnamefont {T.-H.}\ \bibnamefont {Leung}}, \bibinfo {author} {\bibfnamefont {M.}~\bibnamefont {Okano}}, \bibinfo {author} {\bibfnamefont {M.}~\bibnamefont {Block}}, \bibinfo {author} {\bibfnamefont {N.~Y.}\ \bibnamefont {Yao}},\ and\ \bibinfo {author} {\bibfnamefont {D.~M.}\ \bibnamefont {Stamper-Kurn}},\ }\bibfield  {title} {\bibinfo {title} {Spatial coherence of a strongly interacting bose gas in the trimerized kagome lattice},\ }\href@noop {} {\bibfield  {journal} {\bibinfo  {journal} {Physical Review A}\ }\textbf {\bibinfo {volume} {101}} (\bibinfo {year} {2020})}\BibitemShut {NoStop}%
\bibitem [{\citenamefont {Kempkes}\ \emph {et~al.}(2019)\citenamefont {Kempkes}, \citenamefont {Slot}, \citenamefont {van~den Broeke}, \citenamefont {Capiod}, \citenamefont {Benalcazar}, \citenamefont {Vanmaekelbergh}, \citenamefont {Bercioux}, \citenamefont {Swart},\ and\ \citenamefont {Smith}}]{kempkes}%
  \BibitemOpen
  \bibfield  {author} {\bibinfo {author} {\bibfnamefont {S.~N.}\ \bibnamefont {Kempkes}}, \bibinfo {author} {\bibfnamefont {M.~R.}\ \bibnamefont {Slot}}, \bibinfo {author} {\bibfnamefont {J.~J.}\ \bibnamefont {van~den Broeke}}, \bibinfo {author} {\bibfnamefont {P.}~\bibnamefont {Capiod}}, \bibinfo {author} {\bibfnamefont {W.~A.}\ \bibnamefont {Benalcazar}}, \bibinfo {author} {\bibfnamefont {D.}~\bibnamefont {Vanmaekelbergh}}, \bibinfo {author} {\bibfnamefont {D.}~\bibnamefont {Bercioux}}, \bibinfo {author} {\bibfnamefont {I.}~\bibnamefont {Swart}},\ and\ \bibinfo {author} {\bibfnamefont {C.~M.}\ \bibnamefont {Smith}},\ }\bibfield  {title} {\bibinfo {title} {Robust zero-energy modes in an electronic higher-order topological insulator},\ }\href@noop {} {\bibfield  {journal} {\bibinfo  {journal} {Nature Materials}\ }\textbf {\bibinfo {volume} {18}} (\bibinfo {year} {2019})}\BibitemShut {NoStop}%
\bibitem [{\citenamefont {Law}\ \emph {et~al.}(2009)\citenamefont {Law}, \citenamefont {Saxena}, \citenamefont {Kevrekidis},\ and\ \citenamefont {Bishop}}]{kev2}%
  \BibitemOpen
  \bibfield  {author} {\bibinfo {author} {\bibfnamefont {K.~J.~H.}\ \bibnamefont {Law}}, \bibinfo {author} {\bibfnamefont {A.}~\bibnamefont {Saxena}}, \bibinfo {author} {\bibfnamefont {P.~G.}\ \bibnamefont {Kevrekidis}},\ and\ \bibinfo {author} {\bibfnamefont {A.~R.}\ \bibnamefont {Bishop}},\ }\bibfield  {title} {\bibinfo {title} {Localized structures in kagome lattices},\ }\href@noop {} {\bibfield  {journal} {\bibinfo  {journal} {Physical Review A}\ }\textbf {\bibinfo {volume} {79}} (\bibinfo {year} {2009})}\BibitemShut {NoStop}%
\bibitem [{\citenamefont {Vicencio}\ and\ \citenamefont {Johansson}(2013)}]{johansson}%
  \BibitemOpen
  \bibfield  {author} {\bibinfo {author} {\bibfnamefont {R.~A.}\ \bibnamefont {Vicencio}}\ and\ \bibinfo {author} {\bibfnamefont {M.}~\bibnamefont {Johansson}},\ }\bibfield  {title} {\bibinfo {title} {Discrete flat-band solitons in the kagome lattice},\ }\href@noop {} {\bibfield  {journal} {\bibinfo  {journal} {Physical Review A}\ }\textbf {\bibinfo {volume} {87}} (\bibinfo {year} {2013})}\BibitemShut {NoStop}%
\bibitem [{\citenamefont {Kirsch}\ \emph {et~al.}(2021)\citenamefont {Kirsch}, \citenamefont {Zhang}, \citenamefont {Lukas}, \citenamefont {Maczewsky}, \citenamefont {Ivanov}, \citenamefont {Kartashov}, \citenamefont {Torner}, \citenamefont {Bauer}, \citenamefont {Szameit},\ and\ \citenamefont {Heinrich}}]{heinrich}%
  \BibitemOpen
  \bibfield  {author} {\bibinfo {author} {\bibfnamefont {M.~S.}\ \bibnamefont {Kirsch}}, \bibinfo {author} {\bibfnamefont {Y.}~\bibnamefont {Zhang}}, \bibinfo {author} {\bibfnamefont {M.~K.}\ \bibnamefont {Lukas}}, \bibinfo {author} {\bibfnamefont {J.}~\bibnamefont {Maczewsky}}, \bibinfo {author} {\bibfnamefont {S.~K.}\ \bibnamefont {Ivanov}}, \bibinfo {author} {\bibfnamefont {Y.~V.}\ \bibnamefont {Kartashov}}, \bibinfo {author} {\bibfnamefont {L.}~\bibnamefont {Torner}}, \bibinfo {author} {\bibfnamefont {D.}~\bibnamefont {Bauer}}, \bibinfo {author} {\bibfnamefont {A.}~\bibnamefont {Szameit}},\ and\ \bibinfo {author} {\bibfnamefont {M.}~\bibnamefont {Heinrich}},\ }\bibfield  {title} {\bibinfo {title} {Nonlinear second-order photonic topological insulators},\ }\href@noop {} {\bibfield  {journal} {\bibinfo  {journal} {Nature Physics}\ }\textbf {\bibinfo {volume} {17}} (\bibinfo {year} {2021})}\BibitemShut {NoStop}%
\bibitem [{\citenamefont {Prabith}\ \emph {et~al.}(2024)\citenamefont {Prabith}, \citenamefont {Theocharis},\ and\ \citenamefont {Chaunsali}}]{theo}%
  \BibitemOpen
  \bibfield  {author} {\bibinfo {author} {\bibfnamefont {K.}~\bibnamefont {Prabith}}, \bibinfo {author} {\bibfnamefont {G.}~\bibnamefont {Theocharis}},\ and\ \bibinfo {author} {\bibfnamefont {R.}~\bibnamefont {Chaunsali}},\ }\bibfield  {title} {\bibinfo {title} {Nonlinear corner states in topologically nontrivial kagome lattices},\ }\href@noop {} {\bibfield  {journal} {\bibinfo  {journal} {Physical Review B}\ }\textbf {\bibinfo {volume} {110}} (\bibinfo {year} {2024})}\BibitemShut {NoStop}%
\bibitem [{\citenamefont {Hofstrand}\ \emph {et~al.}(2023)\citenamefont {Hofstrand}, \citenamefont {Li},\ and\ \citenamefont {Weinstein}}]{hof}%
  \BibitemOpen
  \bibfield  {author} {\bibinfo {author} {\bibfnamefont {A.}~\bibnamefont {Hofstrand}}, \bibinfo {author} {\bibfnamefont {H.}~\bibnamefont {Li}},\ and\ \bibinfo {author} {\bibfnamefont {M.~I.}\ \bibnamefont {Weinstein}},\ }\bibfield  {title} {\bibinfo {title} {Discrete breathers of nonlinear dimer lattices: bridging the anti-continuous and continuous limits},\ }\href@noop {} {\bibfield  {journal} {\bibinfo  {journal} {Journal of Nonlinear Science}\ }\textbf {\bibinfo {volume} {33}} (\bibinfo {year} {2023})}\BibitemShut {NoStop}%
\bibitem [{\citenamefont {Schindler}\ \emph {et~al.}(2025)\citenamefont {Schindler}, \citenamefont {Bulchandani},\ and\ \citenamefont {Benalcazar}}]{benal}%
  \BibitemOpen
  \bibfield  {author} {\bibinfo {author} {\bibfnamefont {F.}~\bibnamefont {Schindler}}, \bibinfo {author} {\bibfnamefont {V.~B.}\ \bibnamefont {Bulchandani}},\ and\ \bibinfo {author} {\bibfnamefont {W.~A.}\ \bibnamefont {Benalcazar}},\ }\bibfield  {title} {\bibinfo {title} {Nonlinear breathers with crystalline symmetries},\ }\href@noop {} {\bibfield  {journal} {\bibinfo  {journal} {Physical Review B}\ }\textbf {\bibinfo {volume} {111}} (\bibinfo {year} {2025})}\BibitemShut {NoStop}%
\bibitem [{\citenamefont {Flach}(1996)}]{flach3}%
  \BibitemOpen
  \bibfield  {author} {\bibinfo {author} {\bibfnamefont {S.}~\bibnamefont {Flach}},\ }\bibfield  {title} {\bibinfo {title} {Tangent bifurcation of band edge plane waves, dynamical symmetry breaking and vibrational localization},\ }\href@noop {} {\bibfield  {journal} {\bibinfo  {journal} {Physica D}\ }\textbf {\bibinfo {volume} {91}} (\bibinfo {year} {1996})}\BibitemShut {NoStop}%
\bibitem [{\citenamefont {Kapitula}\ and\ \citenamefont {Promislow}(2013)}]{promislow}%
  \BibitemOpen
  \bibfield  {author} {\bibinfo {author} {\bibfnamefont {T.}~\bibnamefont {Kapitula}}\ and\ \bibinfo {author} {\bibfnamefont {K.}~\bibnamefont {Promislow}},\ }\href@noop {} {\emph {\bibinfo {title} {Spectral and Dynamical Stability of Nonlinear Waves}}}\ (\bibinfo  {publisher} {Springer},\ \bibinfo {year} {2013})\BibitemShut {NoStop}%
\bibitem [{\citenamefont {Marin}\ and\ \citenamefont {Aubry}(1998)}]{marin}%
  \BibitemOpen
  \bibfield  {author} {\bibinfo {author} {\bibfnamefont {J.~L.}\ \bibnamefont {Marin}}\ and\ \bibinfo {author} {\bibfnamefont {S.}~\bibnamefont {Aubry}},\ }\bibfield  {title} {\bibinfo {title} {Finite size effects on instabilities of discrete breathers},\ }\href@noop {} {\bibfield  {journal} {\bibinfo  {journal} {Physica D}\ }\textbf {\bibinfo {volume} {119}} (\bibinfo {year} {1998})}\BibitemShut {NoStop}%
\bibitem [{\citenamefont {Fibich}(2015)}]{fibich}%
  \BibitemOpen
  \bibfield  {author} {\bibinfo {author} {\bibfnamefont {G.}~\bibnamefont {Fibich}},\ }\href@noop {} {\emph {\bibinfo {title} {The Nonlinear Schrodinger Equation: Singular Solutions and Optical Collapse}}}\ (\bibinfo  {publisher} {Springer},\ \bibinfo {year} {2015})\BibitemShut {NoStop}%
\bibitem [{\citenamefont {Danieli}\ \emph {et~al.}(2018)\citenamefont {Danieli}, \citenamefont {Maluckov},\ and\ \citenamefont {Flach}}]{flach2}%
  \BibitemOpen
  \bibfield  {author} {\bibinfo {author} {\bibfnamefont {C.}~\bibnamefont {Danieli}}, \bibinfo {author} {\bibfnamefont {A.}~\bibnamefont {Maluckov}},\ and\ \bibinfo {author} {\bibfnamefont {S.}~\bibnamefont {Flach}},\ }\bibfield  {title} {\bibinfo {title} {Compact discrete breathers on flat-band networks},\ }\href@noop {} {\bibfield  {journal} {\bibinfo  {journal} {Low Temp. Phys.}\ }\textbf {\bibinfo {volume} {44}} (\bibinfo {year} {2018})}\BibitemShut {NoStop}%
\bibitem [{\citenamefont {Abramowitz}\ and\ \citenamefont {Stegun}(1980)}]{stegun}%
  \BibitemOpen
  \bibfield  {author} {\bibinfo {author} {\bibfnamefont {M.}~\bibnamefont {Abramowitz}}\ and\ \bibinfo {author} {\bibfnamefont {I.}~\bibnamefont {Stegun}},\ }\href@noop {} {\emph {\bibinfo {title} {Handbook of Mathematical Functions}}}\ (\bibinfo  {publisher} {Dover},\ \bibinfo {year} {1980})\BibitemShut {NoStop}%
\bibitem [{\citenamefont {Feis}\ \emph {et~al.}(2025)\citenamefont {Feis}, \citenamefont {Weidemann}, \citenamefont {Sheppard}, \citenamefont {Price},\ and\ \citenamefont {Szameit}}]{sza}%
  \BibitemOpen
  \bibfield  {author} {\bibinfo {author} {\bibfnamefont {J.}~\bibnamefont {Feis}}, \bibinfo {author} {\bibfnamefont {S.}~\bibnamefont {Weidemann}}, \bibinfo {author} {\bibfnamefont {T.}~\bibnamefont {Sheppard}}, \bibinfo {author} {\bibfnamefont {H.~M.}\ \bibnamefont {Price}},\ and\ \bibinfo {author} {\bibfnamefont {A.}~\bibnamefont {Szameit}},\ }\bibfield  {title} {\bibinfo {title} {Space-time-topological events in photonic quantum walks},\ }\href@noop {} {\bibfield  {journal} {\bibinfo  {journal} {Nature Photonics}\ } (\bibinfo {year} {2025})}\BibitemShut {NoStop}%
\end{thebibliography}%
\clearpage
\appendix
\section{Band structure of the linear lattice}
We look for plane-wave solutions to the linearized equations in \eqref{eq1} in the form
\[A e^{i((n\mathbf{a}_1+m\mathbf{a}_2)\cdot\mathbf{k}-\omega t)}+c.c.\]
Recall that $\mathbf{a}_{1,2}$ are the Kagome lattice vectors and $\mathbf{k}=(k_x,k_y)^{\top}\in\mathbb{R}^2$ is the quasi-momentum.  The vector $A\in\mathbb{C}^3\neq 0$ and $c.c.$ denotes the complex conjugate of the preceding term.  Plugging in the plane-wave ansatz into \eqref{eq1} results in the following $3\times 3$ self-adjoint eigenvalue problem 
\[\omega^2A=\mathcal{M}A,\]
where
\begin{equation}
\mathcal{M}=
\begin{pmatrix}
   \omega_0^2 & -\gamma & -\gamma \\
   -\gamma & \omega_0^2 & -\gamma \\
   -\gamma & -\gamma & \omega_0^2
    \end{pmatrix}
    -\lambda
    \begin{pmatrix}
   0 & e^{i\mathbf{a}_2\cdot\mathbf{k}} & e^{i\mathbf{a}_1\cdot\mathbf{k}} \\
   e^{-i\mathbf{a}_2\cdot\mathbf{k}} & 0 & e^{i\mathbf{a}_1\cdot\mathbf{k}}e^{-i\mathbf{a}_2\cdot\mathbf{k}} \\
   e^{-i\mathbf{a}_1\cdot\mathbf{k}} & e^{-i\mathbf{a}_1\cdot\mathbf{k}}e^{i\mathbf{a}_2\cdot\mathbf{k}} & 0
    \end{pmatrix}
\end{equation}
The real eigenvalues, $\omega_{1,2,3}^2$, are parameterized by the quasi-momentum (taking values over the reciprocal Kagome lattice's hexagonal Brillouin zone $\mathcal{B}$) and the out-of-cell coupling, $\lambda$. They are given explicitly by
\begin{widetext}
\begin{align}\label{disp}
\omega_{1,2}^2(\mathbf{k};\lambda)=\dfrac{1}{2}\left[-\gamma - \lambda + 2\omega_0^2 \pm \sqrt{9\gamma^2 - 6\gamma\lambda + 9\lambda^2 + 
    8\gamma\lambda(\cos(2 k_x) + 2\cos(k_x)\cos(\sqrt{3}k_y))}\right],\enspace  \omega_3^2(\mathbf{k};\lambda)=\gamma + \lambda + \omega_0^2.
\end{align}
\end{widetext}
We remark that $\omega_3^2$ is independent of $\mathbf{k}$. The corresponding eigenvector (not shown) depends on $\mathbf{k}$, but is independent of the parameters $\omega_0$, $\gamma$, and $\lambda$. The labeled points in $\mathcal{B}$ are $\Gamma=(0,0)$, $M=(0,\pi/\sqrt{3})$, $K=(2\pi/3,0)$, and $K'=(-2\pi/3,0).$

\section{Multiple-scale derivation of weakly nonlinear breather envelopes near bottom phonon band}
Here, we outline the derivation of the asymptotic expressions for the breather norms stated in \eqref{norms}.  First, using the parameter $\epsilon=\gamma/2-\lambda$, we re-scale the coordinates, $x_{n,m}^{J}=\epsilon^{1/2\sigma}Y_{n,m}^{J}$, and express system \eqref{eq1} in terms of $\epsilon$ and $Y_{n,m}^{J}$.  Taking the continuum limit\cite{hof}, we introduce $u_{J}(t,\zeta,\eta)\sim Y_{n,m}^{J}(t)$, with continuous spatial variables $\zeta$ and $\eta$. Due to the parabolic behavior of the bottom dispersion band near $\Gamma$, we truncate the resulting Taylor series at second-order, and \eqref{eq1} becomes the following system of PDEs:
\begin{widetext}
\begin{align*}
\partial_t^2u_A =&-\omega_0^2u_A-g\epsilon u_A^{2\sigma+1}+\gamma\left[\dfrac{3}{2}u_B-\dfrac{\sqrt{3}}{2}\partial_{\eta}u_B+\dfrac{3}{4}\partial_{\eta}^2u_B
-\dfrac{1}{2}\partial_{\zeta}u_B+\dfrac{\sqrt{3}}{2}\partial_{\zeta}\partial_{\eta}u_B+\dfrac{1}{4}\partial_{\zeta}^2u_B+\dfrac{3}{2}u_C-\partial_{\zeta}u_C+\partial_{\zeta}^2u_C\right]\\ \nonumber
&-\epsilon\left[u_B-\sqrt{3}\partial_{\eta}u_B+\dfrac{3}{2}\partial_{\eta}^2u_B-\partial_{\zeta}u_B+\sqrt{3}\partial_{\zeta}\partial_{\eta}u_B +\dfrac{1}{2}\partial_{\zeta}^2u_B+u_C-2\partial_{\zeta}u_C+2\partial_{\zeta}^2u_C\right]\\ \nonumber
\partial_t^2u_B =&-\omega_0^2u_B-g\epsilon u_B^{2\sigma+1}+\gamma\left[\dfrac{3}{2}u_A+\dfrac{\sqrt{3}}{2}\partial_{\eta}u_A+\dfrac{3}{4}\partial_{\eta}^2u_A
+\dfrac{1}{2}\partial_{\zeta}u_A+\dfrac{\sqrt{3}}{2}\partial_{\zeta}\partial_{\eta}u_A+\dfrac{1}{4}\partial_{\zeta}^2u_A+\dfrac{3}{2}u_C+\dfrac{\sqrt{3}}{2}\partial_{\eta}u_C
+\dfrac{3}{4}\partial_{\eta}^2u_C\right.\\ \nonumber
&\left.-\dfrac{1}{2}\partial_{\zeta}u_C-\dfrac{\sqrt{3}}{2}\partial_{\zeta}\partial{\eta}u_C+\dfrac{1}{4}\partial_{\zeta}^2u_C\right]
-\epsilon\left[u_A
+\sqrt{3}\partial_{\eta}u_A+\dfrac{3}{2}\partial_{\zeta}^2u_A+\partial_{\zeta}u_A+\sqrt{3}\partial_{\zeta}\partial{\eta}u_A+\dfrac{1}{2}\partial_{\zeta}^2u_A
+u_C+\sqrt{3}\partial_{\eta}u_C\right.\\ \nonumber
&\left.+\dfrac{3}{2}\partial_{\eta}^2u_C-\partial_{\zeta}u_C-\sqrt{3}\partial_{\eta}\partial_{\zeta}u_C+\dfrac{1}{2}\partial_{\zeta}^2u_C\right]\\ \nonumber
\partial_t^2u_C =&-\omega_0^2u_C-g\epsilon u_C^{2\sigma+1}+\gamma\left[\dfrac{3}{2}u_A+\partial_{\zeta}u_A+\partial_{\zeta}^2u_A+\dfrac{3}{2}u_B
-\dfrac{\sqrt{3}}{2}\partial_{\eta}u_B+\dfrac{3}{4}\partial_{\eta}^2u_B
+\dfrac{1}{2}\partial_{\zeta}u_B-\dfrac{\sqrt{3}}{2}\partial_{\zeta}\partial_{\eta}u_B+\dfrac{1}{4}\partial_{\zeta}^2u_B\right]\\ \nonumber
&-\epsilon\left[u_A+2\partial_{\zeta}u_A+2\partial_{\zeta}^2u_A+u_B-\sqrt{3}\partial_{\eta}u_B+\dfrac{3}{2}\partial_{\eta}^2u_B+\partial_{\zeta}u_B-\sqrt{3}\partial_{\zeta}\partial_{\eta}u_B+\dfrac{1}{2}\partial_{\zeta}^2u_B\right].
\end{align*}
\end{widetext}

For $0<\epsilon\ll 1$, we define the long spatiotemporal independent scales: \[Z=\sqrt{\epsilon}\zeta, H=\sqrt{\epsilon}\eta, \text{ and } T=\epsilon t,\]
and expand in powers of $\sqrt{\epsilon}$:
\[u_{J}(t,T,\zeta,Z,\eta,H)=u_{J}^{(0)}+\sqrt{\epsilon}u_{J}^{(1)}+\epsilon u_{J}^{(2)}+\cdots\]  

At \underline{zeroth-order}: \[\mathcal{L}\begin{pmatrix} u_A^{(0)}\\
u_B^{(0)}\\
u_C^{(0)}
    \end{pmatrix}=0,\]
where
\begin{widetext}
\begin{align*}
\hspace{-1.2cm}
\mathcal{L}=&\left(\partial_t^2+\omega_0^2\right)\mathcal{I}_{3\times 3}
+\dfrac{1}{2}\gamma\begin{pmatrix}
0 & \sqrt{3}\partial_{\eta}+\partial_{\zeta}-3 & 2\partial_{\zeta}-3\\
-\sqrt{3}\partial_{\eta}-\partial_{\zeta}-3 & 0 & -\sqrt{3}\partial_{\eta}+\partial_{\zeta}-3\\
-2\partial_{\zeta}-3 & \sqrt{3}\partial_{\eta}-\partial_{\zeta}-3 & 0
\end{pmatrix}
-\dfrac{1}{4}\gamma\begin{pmatrix}
0 & \left(\sqrt{3}\partial_{\eta}+\partial_{\zeta}\right)^2 & 4\partial_{\zeta}^2\\
\left(\sqrt{3}\partial_{\eta}+\partial_{\zeta}\right)^2 & 0 & \left(\sqrt{3}\partial_{\eta}-\partial_{\zeta}\right)^2\\
4\partial_{\zeta}^2 & \left(\sqrt{3}\partial_{\eta}-\partial_{\zeta}\right)^2 & 0
\end{pmatrix},
\end{align*}
\end{widetext}
and where $\mathcal{I}_{3\times3}$ is the identity matrix.

We look for solutions to the zeroth-order system in the form:
\begin{equation*}
u^{(0)}(t,T,\zeta,Z,\eta,H)=R(T,Z,H)ve^{i\omega t}+c.c.
\end{equation*}
for nonzero $v\in\mathbb{C}^3$, where $c.c.$ again denotes complex conjugate of the term preceding it.  Substituting and solving leads to the orthonormal eigensystem:
\begin{align*}
&\omega_{-}^2:=\omega_0^2-3\gamma\>\text{ and }\> v_-:=\dfrac{1}{\sqrt{3}}\begin{pmatrix} 1\\ 1\\ 1\end{pmatrix}\\
&\omega_{+}^2:=\omega_0^2+\dfrac{3}{2}\gamma\>\text{ and }\> v_{+,1\setminus 2}:=\dfrac{1}{\sqrt{2}}\begin{pmatrix} \text{-}1\\ 0\\ 1\end{pmatrix},\> \dfrac{1}{\sqrt{2}}\begin{pmatrix} \text{-}1\\ 1\\ 0\end{pmatrix}.
\end{align*}
Here, we are interested in breathers with fixed frequency $\omega_b=\omega_-$ and so consider zeroth-order solutions in the form
\begin{equation*}
u^{(0)}=R(T,Z,H)\sqrt{3}v_-e^{i\omega_-t}+c.c.
\end{equation*}

At \underline{first-order}:
\begin{equation*}\mathcal{L}\begin{pmatrix} u_A^{(1)}\\
u_B^{(1)}\\
u_C^{(1)}
    \end{pmatrix}=\gamma\begin{pmatrix}
   -\dfrac{3}{2}\partial_Z R -\dfrac{\sqrt{3}}{2}\partial_H R\\
   \sqrt{3}\partial_H R \\
   \dfrac{3}{2}\partial_ZR-\dfrac{\sqrt{3}}{2}\partial_{H} R
    \end{pmatrix}e^{i\omega_-t}+c.c.
\end{equation*}

The right-hand side of the above equation can be expressed 
\[\dfrac{\sqrt{2}}{2}\gamma\left(3v_{+,1}\partial_Z R+\sqrt{3}\left(2v_{+,2}-v_{+,1}\right)\partial_H R\right)e^{i\omega_-t}+c.c.\]
Since the operator $\mathcal{L}$ is self-adjoint and the expression above is contained in the orthogonal complement of $\mathcal{L}$'s kernel, the Fredholm alternative implies the system's solvability.  Indeed, a particular solution, $u_p^{(1)}$, of the first-order system is given by a constant multiple of the above expression:
\[u_p^{(1)}=\dfrac{\sqrt{2}}{3}\left(v_{+,1}\partial_Z R+\dfrac{1}{\sqrt{3}}\left(2v_{+,2}-v_{+,1}\right)\partial_H R\right)e^{i\omega_-t}+c.c.\]

Plugging in the zeroth and first order solutions gives at 
\underline{second-order}:
\vspace{0.2cm}
\begin{widetext}
\begin{align*}
\mathcal{L}\begin{pmatrix} u_A^{(2)}\\
u_B^{(2)}\\
u_C^{(2)}
    \end{pmatrix}=&\left[\left(-2i\omega_-\partial_TR-g\binom{2\sigma+1}{\sigma}|R|^{2\sigma}R-2R\right)\sqrt{3}v_{-}
    +\dfrac{\gamma}{12}\partial_Z^2R\begin{pmatrix} 11\\
    2\\
    11
    \end{pmatrix}-\dfrac{\sqrt{6}\gamma}{2}\partial_Z\partial_HRv_{+,1}+\dfrac{\gamma}{12}\partial_H^2R\begin{pmatrix}
        5\\
        14\\
        5
    \end{pmatrix}\right]e^{i\omega_-t}\\
    &+\text{ non-resonant terms }+c.c.,
\end{align*}
\end{widetext}
where the coefficient to the nonlinear term is a binomial coefficient.

By taking the orthogonal projection of the above right-hand-side onto $v_-$ and forcing the resonant terms to vanish leads to the following NLS equation:
\begin{equation}\label{nls}
-2i\omega_-\partial_TR-g\binom{2\sigma+1}{\sigma}|R|^{2\sigma}R-2R+\dfrac{2}{3}\gamma\Delta R=0
\end{equation}

For a softening nonlinearity, $g=-1$, the above generalized NLS equation is of ``focusing" type and has spatially localized solitary wave solutions.  We seek solutions of the form
\[R(T,Z,H)=S(Z,H; \nu)e^{i\nu T/2\omega_-}\]
which gives the asymptotic solutions on the BK lattice:
\begin{equation}
\begin{pmatrix}
x_{n,m}^A(t)\\
x_{n,m}^B(t)\\
x_{n,m}^C(t)
    \end{pmatrix}\sim 2\epsilon^{1/2\sigma}S(\sqrt{\epsilon}n,\sqrt{\epsilon }m;\nu)\begin{pmatrix} 1\\ 1\\ 1\end{pmatrix}\cos\left(\left[\omega_-+\dfrac{\nu\epsilon}{2\omega_-}\right]t\right).
    \label{asymp}
\end{equation}

We can then use \eqref{asymp} to compute the spatial norms of the small-amplitude breathers stated in \eqref{norms}.  The constant term in the second expression in \eqref{norms} is due to the area of the hexagonal Voronoi cell of the Kagome lattice. In spatial dimensions greater than one, the ground-state radial solitary wave solution of the focusing NLS, $S_0$, is not given explicitly, but can be found numerically. Here we use the \textit{spectral renormalization method}\cite{fibich} (In Figure \ref{figure4} we use $\|S_0\|\approx 1.61$ for $\sigma=1$ and $\|S_0\|\approx 0.77$ for $\sigma=2$).

\end{document}